\documentclass[pre,twocolumn]{revtex4}
\usepackage[english]{babel}
\usepackage{amsmath}
\usepackage{amssymb}
\usepackage{epsfig}

\begin{document}
\title{Direct and reverse precession of a massive vortex\\
in a binary Bose--Einstein condensate}
\author{Victor P. Ruban}
\email{ruban@itp.ac.ru}
\affiliation{Landau Institute for Theoretical Physics RAS,
Chernogolovka, Moscow region, 142432 Russia}

\date{\today}

\begin{abstract}
The dynamics of a filled massive vortex is studied numerically 
and analytically using a two-dimensional model of a two-component 
Bose--Einstein condensate trapped in a harmonic trap. 
This condensate exhibits phase separation. In the framework of the 
coupled Gross--Pitaevskii equations, it is demonstrated that, 
in a certain range of parameters of the nonlinear interaction, 
the precession of a sufficiently massive vortex around the center 
is strongly slowed down and even reverses its direction with a further 
increase in the mass. An approximate ordinary differential equation 
is derived that makes it possible to explain this behavior of the system.
\end{abstract}

\maketitle

\section*{Introduction}

Ultracold gas mixtures consisting either of different
chemical (alkaline) elements, or of different isotopes
of the same element, or of the same isotopes in two
different (hyperfine) quantum states exhibit a much
wider variety of static and dynamic properties as compared 
to single-component Bose--Einstein condensates [1--5]. 
To a large extent, this is due to the presence 
of several parameters of nonlinear interactions,
which are proportional to the corresponding scattering
lengths and can sometimes be tuned over a wide range
according to the aims of experimentalists using Feshbach 
resonances [6--10]. In particular, with a sufficiently 
strong mutual repulsion between two types of matter waves, 
a regime involving spatial separation of condensates is possible [11,12], 
in which domain walls are formed between the phases, characterized 
by effective surface tension [4,13]. Spatial separation can
underlie many interesting configurations and phenomena, 
for example, the nontrivial geometry of the ground state 
of binary immiscible Bose–Einstein condensates in traps [14--16] 
(including optical lattices [17--19]), bubble dynamics [20], 
quantum analogs of classical hydrodynamic instabilities 
(Kelvin--Helmholtz [21,22], Rayleigh--Taylor [23--25], 
and Plateau--Rayleigh [26]), parametric instability of capillary 
waves at the interface [27,28], complex textures in
rotating binary condensates [29--31], three-dimensional 
topological structures [32–37], capillary buoyancy of dense droplets 
in trapped immiscible Bose--Einstein condensates [38], etc.

In particular, vortices with a filled core in binary
Bose--Einstein condensates and their dynamics are of
considerable interest [3,39--45]. Such a structure can
be represented as a quantized vortex in one of the components, 
the core of which is filled with the second component 
(see a numerical example in Fig.1). In this case, 
the dip in the density of the vortex component 
provides a potential well trapping the second
(``bright'') component. In turn, the bright component
creates a potential ``hill that pushes'' the vortex component 
apart and thereby increases the width of the vortex core. 
As a result, some equilibrium profile is formed in a self-consistent manner.

As compared to vortices in the $B$-phase of superfluid ${}^3$He
occupied by the chiral $A$-phase, filled vortices
in ultracold rarefied mixtures of Bose gases have a
much simpler structure (for comparison, see review
[46] and also [47]). In particular, in the core of a (stationary) 
cold gas vortex, the superfluid current has a simple structure
${\bf j}\propto[\rho_1(r)/r]{\bf e}_\varphi$ (in contrast to 
Fig.41 in [46], where there exists a counter-rotation region).
Moreover, since binary Bose--Einstein condensates of
cold atoms are described by a system of coupled
Gross--Pitaevskii equations, in which the Hamiltonian 
contains no cross terms in the kinetic energy,
there is no well-known Andreev--Bashkin effect,
where the superfluid velocity of one component contributes 
to the current of another component [48, 49].

The vortex complex in the external potential of the
trap, being deviated from the equilibrium position,
begins to move nearly as a whole. One of the aims of
researchers is the theoretical analysis of the emerging
motion and the derivation of effective simplified equations 
for its description. For example, spatially two-dimensional 
models of binary condensates in a potential 
well with a flat bottom were recently considered in
[42,43], where ordinary differential equations were
proposed to determine the dynamics of massive
``pointlike'' vortices in such systems. At the same time,
the practically important case of a smooth external
potential remains unexplored. In this work, this gap is
filled. Approximate equations of motion for massive
vortices in smoothly inhomogeneous Bose--Einstein
condensates are derived. These equations have a non-canonical 
Hamiltonian structure and, for one vortex,
formally correspond to the dynamics of an electric
charge in a certain static (two-dimensional) inhomogeneous 
electromagnetic field, where the electrostatic
field lies in the $(x,y)$ plane, and the magnetic field is
directed along the $z$ axis and is proportional to the
equilibrium density $\rho(x,y)$ of the condensate without vortices.

\begin{figure}
\begin{center}
(a)\epsfig{file=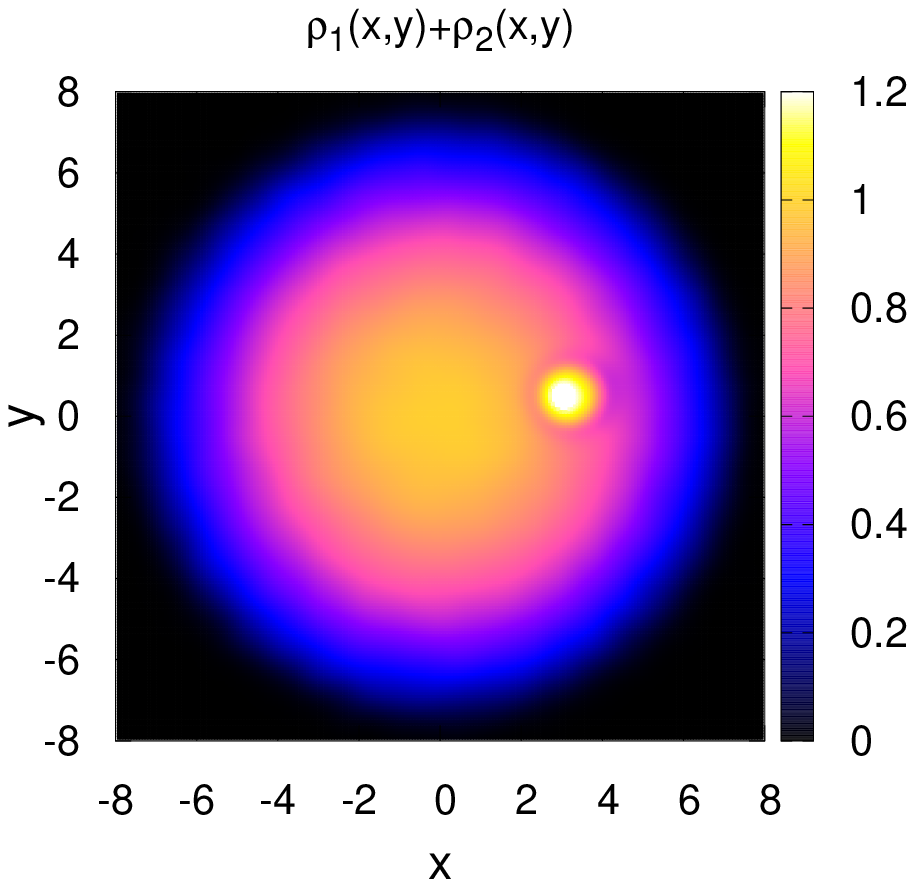, width=84mm}\\
(b)\epsfig{file=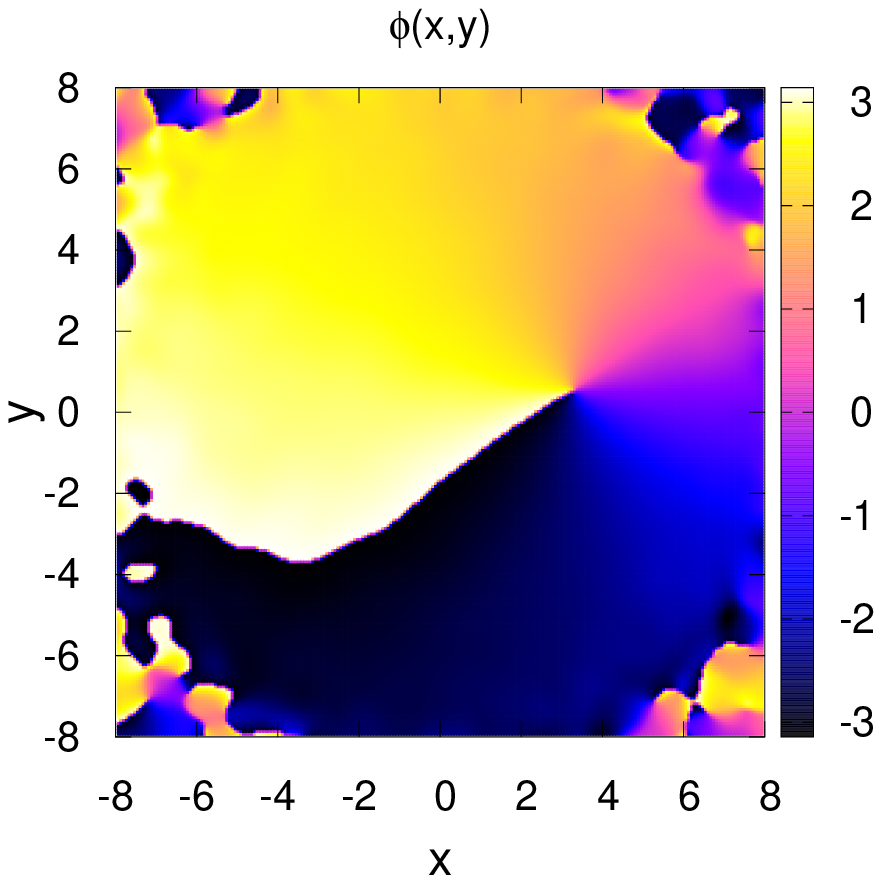, width=84mm}
\end{center}
\caption{
Numerical example of a filled vortex: (a) normalized total density 
of condensates and (b) the phase of the vortex component.
}
\label{VB} 
\end{figure}

Analyzing the dependence of the parameters of this
simplified model on the parameters of the original system 
of partial differential equations (in our case, these
are coupled two-dimensional Gross--Pitaevskii equations), 
we reveal an interesting effect that has not yet
been recognized for vortices in Bose--Einstein condensates. 
Namely, in the case of unequal nonlinear self-repulsion 
coefficients, the ``electric'' force in the case of 
a small filling of vortices with the bright component 
(having small effective mass) can be directed
from the center of the system, and as the mass
increases, it gradually decreases and then reverses its
direction. This leads to a change in the direction of the
vortex drift (its precession around the origin). We
emphasize that, with such reverse precession, no qualitative 
rearrangement occurs inside the vortex core.
Direct numerical simulations of the Gross--Pitaevskii
equations will confirm the predictions of the simplified model.

\section*{Initial model}

We consider a two-dimensional, sufficiently rarefied 
binary Bose--Einstein condensate in the limit of
zero temperature, where the Gross--Pitaevskii equations 
are applicable. For maximum simplicity, it is
assumed that both types of atoms have the same mass,
$m_1=m_2=m$. Under this assumption, the case of a
small difference in the masses of isotopes, such as $^{85}$Rb
and $^{87}$Rb, can be described approximately. The harmonic 
trap is characterized by a transverse frequency $\omega$, 
which is the same for both types of atoms. We
choose the scale $\tau=1/\omega$ for time, $l_{\rm tr}=\sqrt{\hbar/\omega m}$
for length, and $\varepsilon=\hbar\omega$ for energy. This allows us to write
the equations of motion for the complex wavefunctions $A({\bf r},t)$
(vortex component) and $B({\bf r},t)$ (bright component) 
in the dimensionless form
\begin{eqnarray}
i\dot A=-\frac{1}{2}\nabla^2 A+\left[V(x,y)+g_{11}|A|^2+g_{12}|B|^2\right]A,&&
\label{GP1}\\
i\dot B=-\frac{1}{2}\nabla^2 B+\left[V(x,y)+g_{21}|A|^2+g_{22}|B|^2\right]B,&&
\label{GP2}
\end{eqnarray}
where $V=(x^2+y^2)/2$ is the trap potential and $g_{\alpha\beta}$ is
the symmetric matrix of nonlinear interactions. The
interactions can be physically described by the scattering 
lengths $a_{\alpha\beta}$ [2]:
\begin{equation}
g^{\rm phys}_{\alpha\beta}=
2\pi \hbar^2 a_{\alpha\beta}(m_\alpha^{-1}+m_\beta^{-1}).
\end{equation}
Since we are interested in the situation where all scattering 
lengths are positive, the first self-repulsion factor 
can be normalized to unity, $g_{11}=1$. For only one
component $A$ without any vortices, the equilibrium
condensate density would be
\begin{equation}
|A_0|^2=\rho({\bf r})\approx [\mu-V(\bf r)],
\end{equation}
where $\mu\gg 1$ is the chemical potential. The effective
radius of the condensate is $R_*=\sqrt{2\mu}$. The motion of
the massive vortex complex occurs against such inhomogeneous background.

The phase separation condition $g=(g^2_{12}-g_{11}g_{22})>0$
favors the existence of a filled vortex [11,12]. 
In a relatively narrow transient layer between the 
separated condensates, the densities of both phases nearly 
vanish in one or the opposite direction. However,
for the applicability of the Gross--Pitaevskii equations, 
the width of this layer $\delta \sim 1/\sqrt{g\rho}$ should nevertheless 
be much larger than the characteristic scattering 
length $a$ (usually equal to several hundred Bohr
radii), i.e., $\delta (l_{\rm tr}/a)\gg 1$. The corresponding energy
excess (surface tension) is given by the expression
\begin{equation}
\sigma=K(g_{22}/g_{11},g_{12}/g_{11}) \rho^{3/2},
\end{equation}
where $K\sim \sqrt{g}$ at small $g$ [11,13]. Below, we will see
that the dependence of the surface tension on the
background density significantly affects the dynamics
of a massive vortex, since it creates a gradient of its
effective potential energy.

\section*{Structure of wavefunctions}

When deriving the equations of vortex motion, we
neglect the free excitations of acoustic vibrations and
the deviation of the vortex shape from the circular one.
The decisive factor allowing this to be done is the
smallness of the ratio $w/ R_*$, where $w$ is the radius of
the vortex core and $R_*$ is the size of the condensate.
Under this condition, the structure of the vortex is
almost the same as that against the homogeneous
background and its velocity is much lower than the
velocities of potential excitations. As a result, the
wavefunctions $A$ and $B$ can be approximately represented 
in the simplified form
\begin{eqnarray}
&&A=\sqrt{\rho_a}e^{i\Phi_a}\approx
\Psi_v({\bf r}, {\bf R}) \exp[i{\bf U}\cdot{\bf F}({\bf r},{\bf R})],
\label{A_appr}
\\
&&B=\sqrt{\rho_b}e^{i\Phi_b}\approx
\Psi_b({\bf r}, {\bf R}) \exp[i{\bf U}\cdot{\bf G}({\bf r},{\bf R})],
\label{B_appr}
\end{eqnarray}
where ${\bf R}(t)$ is the vortex position and ${\bf U}(t)$ is a certain
two-dimensional vector, the relation of which to the
vortex velocity will be revealed further on. It is very
important that the function $\Psi_v$ involves a quantized
vortex, and its phase $\Phi_v({\bf r})$ increases by $2\pi$ 
at the counterclockwise passing around the ${\bf R}$ point. The vortex
phase $\Phi_v({\bf r})$ is appropriately matched to the density $\rho_a$,
which coincides with $\rho({\bf r})$ far from the vortex, whereas
$\rho_a\approx \varrho(|{\bf r}-{\bf R}|,{\bf R})$ inside its core, so that
\begin{equation}
\mbox{div} (\rho_a\nabla \Phi_v)=0.
\end{equation}
In turn, the densities $\varrho(|{\bf r}-{\bf R}|,{\bf R})$ and $\rho_b$ 
are matched to each other, as in the case of the filled vortex, against
the locally homogeneous background density 
$\varrho(\infty,{\bf R})=\rho({\bf R})$. Additional functions ${\bf F}$
and ${\bf G}$ in the $A$ and $B$ phases are related to the parts 
of velocity fields that are due to the motion of the dip in the vortex 
component density and to the quasihomogeneous motion
of the bright component, respectively. With a sufficient
accuracy, we can assume that ${\bf G}={\bf r}-{\bf R}$ and 
${\bf F}=({\bf r}-{\bf R})f(|{\bf r}-{\bf R}|, {\bf R})$,
where the scalar function $f(\chi)$ at fixed ${\bf R}$
satisfies the second order linear differential equation
\begin{equation}
\varrho(\chi)[\chi f''+3f']+\varrho'(\chi)[\chi f'+f]=\varrho'(\chi),
\label{f_eq}
\end{equation}
which follows from the continuity equation for the steady-state flow
\begin{equation}
\mbox{div}(\varrho[U_i F_{i,k}-U_k])=0.
\end{equation}
Equation (9) is supplemented by the conditions that $f$
tends to zero at infinity and that the singularity at zero
should be as weak as possible. It is very important that $f$
decreases rapidly at distances of the order of the vortex 
core width. The product ${\bf U}\cdot{\bf F}$ is qualitatively similar 
to the velocity potential that describes the flow
around a cylinder in classical hydrodynamics and
determines the corresponding added mass.

\section*{Simplified equations of motion for a vortex}

We derive approximate equations of motion for the
filled vortex based on the Hamiltonian structure of the
Gross--Pitaevskii equations,
\begin{equation}
i\dot A=\delta{\cal H}/\delta A^*, \qquad i\dot B=\delta{\cal H}/\delta B^*,
\label{canon}
\end{equation}
where the Hamiltonian is given by the expression
\begin{eqnarray}
{\cal H}&=&\int\Big[\frac{|\nabla A|^2}{2}+\frac{|\nabla B|^2}{2}
+V(x,y)(|A|^2+|B|^2)\nonumber\\
&&+\frac{g_{11}|A|^4}{2}+g_{12}|A|^2|B|^2+\frac{g_{22}|B|^4}{2}\Big]dx dy.
\end{eqnarray}

Using Eqs. (6), (7), and (11), we easily obtain the two relations
\begin{eqnarray}
\int \Bigg(\dot\rho_a \frac{\partial \Phi_a}{\partial{\bf R}}
-\dot\Phi_a\frac{\partial \rho_a}{\partial{\bf R}}\Bigg)dx dy + \{a\to b\}
=\frac{\partial {\cal H}}{\partial{\bf R}},&&
\label{deltaR}
\\
\int \Bigg(\dot\rho_a \frac{\partial \Phi_a}{\partial{\bf U}}
-\dot\Phi_a\frac{\partial \rho_a}{\partial{\bf U}}\Bigg)dx dy + \{a\to b\}
=\frac{\partial {\cal H}}{\partial{\bf U}}.&&
\label{deltaU}
\end{eqnarray}
Far from the vortex, either $\dot\rho_{a,b}$ or variational derivatives 
$\partial \rho_{a,b}/\partial{\bf R}$ are negligibly small 
(and $\partial \rho_{a,b}/\partial{\bf U}=0$ everywhere). 
Therefore, the main contribution to the
integrals comes from the closest vicinity of the vortex,
where we can use the approximate formulas
\begin{eqnarray}
\dot A&\approx& -\dot{\bf R}\cdot\nabla A +iA(\dot{\bf U}\cdot {\bf F}),\\
\dot B&\approx& -\dot{\bf R}\cdot\nabla B +iB(\dot{\bf U}\cdot {\bf G}).
\end{eqnarray}
Substituting the time derivatives of densities and
phases expressed from these formulas into Eqs. (13) and (14), 
we obtain after straightforward calculations two vector equations
\begin{eqnarray}
-M\dot{\bf U}+2\pi[\hat z \times \dot{\bf R}]\rho({\bf R})
\approx\partial H/\partial{\bf R},&&
\label{U_t}
\\
M\dot{\bf R}\approx\partial H/\partial{\bf U},&&
\label{R_t}
\end{eqnarray}
where the scalar variable M is determined by the expression
\begin{equation}
M\delta_{ik}=\int(\rho_a F_{k,i}+\rho_b G_{k,i} )dx dy.
\end{equation}
It is clear that the Hamiltonian $H({\bf R},{\bf U})$ is quadratic in
terms of  ${\bf U}$ and should have the form
\begin{equation}
H\approx M_{tot}\frac{{\bf U}^2}{2}+W({\bf R}).
\label{H_RU}
\end{equation}
Therefore, one should identify ${\bf U}$ with the velocity of the vortex
$\dot{\bf R}$, and then $M$ turns out to be the total
effective mass of the filled vortex, which includes both
the mass of the bright component $M_{br}$ trapped in the
vortex core and the added mass of the vortex component 
due to the presence of additional kinetic energy
during the motion of the density dip through the condensate.

In principle, the added mass $M_{add}$ could depend on ${\bf R}$; 
then, to preserve the conservative character of the system, 
we should introduce the vector ${\bf P}=M({\bf R}){\bf U}$
and rewrite the equations of motion in the ``slightly corrected'' form
\begin{eqnarray}
-\dot{\bf P}+2\pi[\hat z \times \dot{\bf R}]\rho({\bf R})
=\partial H/\partial{\bf R},&&
\label{P_t_conserv}
\\
\dot{\bf R}=\partial H/\partial{\bf P}.&&
\label{R_t_conserv}
\end{eqnarray}
Probably, the generalization would be useful when considering 
strongly oblate three-dimensional condensates in the Thomas--Fermi 
regime along all three coordinates. However, it is easy to show that, 
in the case of strictly two-dimensional Gross--Pitaevskii
equations, the mass $M_{add}$ in the leading approximation
does not depend on the position of the vortex.

\section*{Estimates for the coefficients}

If the effective radius $w$ of the vortex significantly
exceeds the thickness of the domain wall $\delta$, then the
added mass can be estimated using the well-known
formula from classical hydrodynamics, i.e., 
$M_{add}\approx \pi\rho({\bf R})w^2$. On the other hand, 
since the hydrodynamic pressures of the condensates are 
$P_a=g_{11}\rho_a^2/2$ and $P_b=g_{22}\rho_b^2/2$, and they coincide 
in the leading order on both sides of the domain wall (if the surface 
tension $\sigma$ is not taken into account), the conservation of the
mass of the bright component makes it possible to estimate 
the dependence $w({\bf R})$ using the condition
\begin{equation}
M_{br}=\pi\rho_b w^2=\pi\sqrt{\frac{g_{11}}{g_{22}}}\rho({\bf R})w^2
=\sqrt{\frac{g_{11}}{g_{22}}} M_{add}.
\end{equation}
In addition, this gives us also the estimate for the effective mass
\begin{equation}
M=[1+(g_{22}/g_{11})^{1/2}]M_{br}.
\end{equation}

Since the total mass of a sufficiently large vortex
turns out to be independent of its position on a spatially
inhomogeneous background density, the derived system 
of equations (17) and (18) [taking into account Eq.(20)] 
is mathematically identical to the equation of
the two-dimensional motion of an electric charge in
magnetic and electric fields. The ``electrostatic potential'' 
$W({\bf R})$ in this case is equal to the sum of three contributions. 
The first contribution is the part of the kinetic energy that 
is due to the gradient of the vortex phase $\Phi_v({\bf r})$. 
The second contribution is the energy of nonlinear interactions 
plus the potential energy of the vortex in the field of the trap. 
The third contribution is the energy of ``quantum pressure.'' 
The first contribution in the well-known local induction approximation
can be estimated as
\begin{equation}
W_1\approx \pi\Lambda_0 \rho({\bf R}), \qquad \Lambda_0=\ln(R_*/w_0).
\end{equation}
Note that $\Lambda_0\approx 2$ in practically interesting cases. The
second and third contributions can be estimated (up to an insignificant
additive constant) using the concept of surface tension,
\begin{eqnarray}
W_2+W_3\approx 2\pi\sigma w + [1-(g_{22}/g_{11})^{1/2}] M_{br} U({\bf R})&&
\nonumber\\
=[\pi C M_{br}^{1/2}-[1-(g_{22}/g_{11})^{1/2}] M_{br}]\rho({\bf R}),&& 
\end{eqnarray}
where $C$  is a coefficient about unity. Thus, the whole
expression for the potential energy is reduced to the simple form
\begin{equation}
W({\bf R})=\pi\Lambda \rho({\bf R}),
\end{equation}
where the effective dimensionless parameter $\Lambda$
depends in a nontrivial way on the mass of the trapped bright component,
\begin{equation}
\Lambda \approx \Lambda_0 + C M_{br}^{1/2}-[1-(g_{22}/g_{11})^{1/2}] M_{br}/\pi.
\label{Lambda}
\end{equation}
Here, the second term is due to the surface tension at
the phase boundary, and the third one is responsible for the 
``mass'' contribution, taking into account the buoyant force. 
It is noteworthy that the hydrodynamic contribution $\Lambda_0$ 
can be significantly smaller than each of the other terms.

\section*{Vortex precession rate}

In terms of the complex position $Z(t)=X(t)+iY(t)$,
of the vortex, the resulting equation 
of motion for the massive vortex can be written as
\begin{equation}
-\tilde M \ddot Z + i(\mu-|Z|^2/2)\dot Z + \frac{\Lambda}{2} Z=0,
\label{ODEv1}
\end{equation}
where $\tilde M=M/2\pi$. This equation is integrable in
quadratures in an obvious way when passing to polar
coordinates. We do not present the corresponding formulas 
here but note only that particular solutions in the form 
$Z(t)=R_0\exp(-i\Omega t)$ exist. The corresponding 
substitution gives two branches of the solution
\begin{equation}
\Omega_\pm(R_0)=-\Big[\frac{(\mu-R_0^2/2)}{2\tilde M} \pm 
\sqrt{\frac{(\mu-R_0^2/2)^2-2\Lambda\tilde M}{4\tilde M^2}}\Big].
\end{equation}
Here, the solution $\Omega_-(R_0)$, which corresponds to the
charge drift in crossed magnetic and electric fields, is
mainly of interest. It is seen that large positive values
of $\Lambda$ promote a fast counterclockwise drift [negative
frequency $\Omega_-(R_0)$]. Moreover, both branches merge at
some critical value of the radius, which decreases with
increasing $\Lambda$. On the contrary, the drift is slowed down
at small $\Lambda$ and reverses its direction (clockwise drift at
positive frequency) at negative $\Lambda$. According to Eq.(28), 
the system under study has two features at $g_{22}\neq g_{11}=1$. 
First, the parameter $\Lambda$ at $g_{22} > 1$ increases
rapidly with the vortex mass, which should reduce the
area of the stability of motion. Second, negative values
of $\Lambda$ and reverse drift are possible at a sufficiently large
vortex mass if $g_{22} < 1$.

\begin{figure}
\begin{center}
\epsfig{file=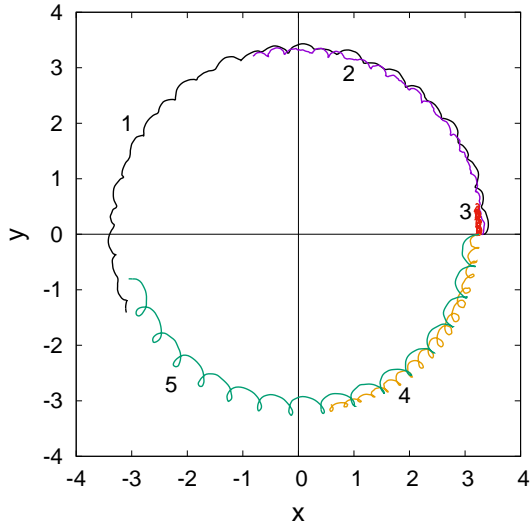, width=80mm}
\end{center}
\caption{
Examples of trajectories of a massive
vortex in the time interval $[0:100]$ at $\mu=30$ for masses of
the bright component $M_{br}/2\pi=$ (1) 8.0, (2) 10.0, (3) 12.0,
(4) 14.0, and (5) 16.0. The motion begins at approximately
the same point, but the direction and velocities of the drift
are different.}
\label{trajectories} 
\end{figure}

\begin{figure}
\begin{center}
\epsfig{file=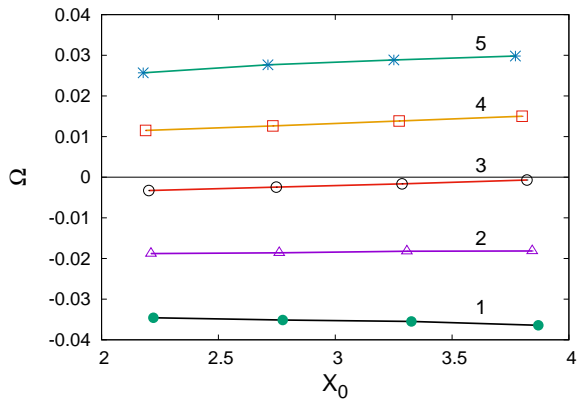, width=80mm}
\end{center}
\caption{
Numerically calculated drift angular velocity versus 
the initial coordinate at $g_{11}=g_{12}=1$, $g_{22}=0.5$, and $\mu=30$ 
for masses of the bright component $M_{br}/2\pi=$ 
(1) 8.0, (2) 10.0, (3) 12.0, (4) 14.0, and (5) 16.0.
}
\label{Omega} 
\end{figure}

\section*{Results of numerical tests}

The parametric domain $\Lambda\gg 1$ will be studied elsewhere. 
Here, the prediction of the theory about the reverse drift 
is verified by direct numerical simulation of the coupled 
two-dimensional Gross--Pitaevskii equations. The performed 
calculations were focused on experimentally realized 
$^{85}$Rb-$^{87}$Rb mixtures [8], where $a_{12}/a_{22}\approx 2$, 
while $a_{11}$ can be varied over a wide range using the Feshbach 
resonance. For this reason, in our numerical experiments, we took 
the values $g_{11}=g_{12}=1$, and $g_{22}=0.5$. The employed numerical 
method is similar to that used in [37,38,50].

The initial state for numerical integration was prepared 
in such a way that the filled vortex was located at
some point ${\bf R}(0)=(X_0,0)$ and had zero velocity. Next,
its trajectory (the position of the center of mass of the
bright component) was tracked and the drift angular
velocity was calculated. Examples of the trajectories
are shown in Fig.2. It is seen that fast oscillations
characteristic of the charge motion in a magnetic field
are superimposed on the slow drift. The plots of the
drift angular velocity, confirming the possibility of the
reverse motion, are shown in Fig.3.

\section*{Conclusions}

To summarize, a simple equation of motion has been derived 
for a massive vortex in a two-dimensional smoothly inhomogeneous 
binary Bose--Einstein condensate and the possibility 
of reverse drift predicted by this equation has been 
numerically confirmed. The equation differs from a similar 
equation for a massless vortex in the natural additional term $M\ddot{\bf R}$
and in a different coefficient in the expression for the effective
potential energy. There are even more possibilities for
controlling the coefficients in the case of unequal
atomic masses $m_1\neq m_2$ and at different external potentials 
$V_1({\bf r})\neq V_2({\bf r})$.

Note that a system of several massive vortices can
be treated in a similar way. Technical difficulty
appears only in the calculation of pair interactions.
However, for some special equilibrium density profiles
$\rho(x,y)$, this difficulty is successfully overcome, as
shown in [51,52] for one-component Bose--Einstein condensates.

A strictly two-dimensional condensate is an idealization. 
A detailed test of the reverse precession regime
for strongly oblate three-dimensional condensates is a
task for the future. Preliminary numerical experiments
have already shown a qualitative similarity to the planar case.

\section*{Conflict of interest}

The author declares that he has no conflicts of interest.

\end{document}